\documentclass[11pt,a4paper]{article}
\pdfoutput=1
\usepackage{jcappub}

\usepackage{color}
\usepackage{ulem}
\usepackage{graphicx}
\usepackage{float}
\usepackage[caption=false]{subfig}
\usepackage{braket}
\usepackage{slashed}

\usepackage{tabularx}
\usepackage[colorlinks=true,citecolor=blue,linkcolor=blue]{hyperref}%
\begin{document}
%

\title{Connecting Dark Matter Annihilation to the Vertex Functions of Standard Model Fermions}

\author[a]{Jason Kumar}
\author[a]{Christopher Light}

\affiliation[a]{\mbox{Department of Physics \& Astronomy, University of Hawaii, Honolulu, HI 96822, USA.}}

\emailAdd{jkumar@hawaii.edu}
\emailAdd{lightc@hawaii.edu}

\def\be{\begin{equation}}
\def\ee{\end{equation}}
\def\al{\alpha}
\def\bea{\begin{eqnarray}}
\def\eea{\end{eqnarray}}

\def\tev{\, {\rm TeV}}
\def\gev{\, {\rm GeV}}
\def\mev{\, {\rm MeV}}
\def\kev{\, {\rm keV}}
\def\swsq{\sin^2\theta_W}
\def\xfb{\, {\rm fb}}
\newcommand{\sigmaSI}{\sigma_{\rm SI}}
\newcommand{\sigmaSD}{\sigma_{\rm SD}}
\newcommand{\gsim}{\lower.7ex\hbox{$\;\stackrel{\textstyle>}{\sim}\;$}}
\newcommand{\lsim}{\lower.7ex\hbox{$\;\stackrel{\textstyle<}{\sim}\;$}}
\newcommand{\fb}{\rm fb}
\newcommand{\ifb}{\rm fb^{-1}}
\newcommand{\pb}{\rm pb}
\newcommand{\ipb}{\rm pb^{-1}}
\newcommand{\m}{\rm m}

\newcommand{\drawsquare}[2]{\hbox{%
\rule{#2pt}{#1pt}\hskip-#2pt
\rule{#1pt}{#2pt}\hskip-#1pt
\rule[#1pt]{#1pt}{#2pt}}\rule[#1pt]{#2pt}{#2pt}\hskip-#2pt
\rule{#2pt}{#1pt}}

\newcommand{\fund}{\raisebox{-.5pt}{\drawsquare{6.5}{0.4}}}
\newcommand{\Ysymm}{\raisebox{-.5pt}{\drawsquare{6.5}{0.4}}\hskip-0.4pt%
        \raisebox{-.5pt}{\drawsquare{6.5}{0.4}}}
\newcommand{\Yasymm}{\raisebox{-3.5pt}{\drawsquare{6.5}{0.4}}\hskip-6.9pt%
        \raisebox{3pt}{\drawsquare{6.5}{0.4}}}
\newcommand{\antifund}{\overline{\fund}}
\newcommand{\bYasymm}{\overline{\Yasymm}}
\newcommand{\bYsymm}{\overline{\Ysymm}}
\newcommand{\Dsl}[1]{\slash\hskip -0.20 cm #1}

\newcommand{\ssection}[1]{{\em #1.\ }}
\newcommand{\Dsle}[1]{\slash\hskip -0.28 cm #1}
\newcommand{\met}{{\Dsle E_T}}
\newcommand{\Dslp}[1]{\slash\hskip -0.23 cm #1}
\newcommand{\mpt}{{\Dslp p_T}}

\abstract{
We consider scenarios in which dark matter is a Majorana fermion which couples to Standard Model
fermions through the exchange of charged mediating particles.  The matrix elements for various dark matter
annihilation processes are then related to one-loop corrections to the fermion-photon vertex, where dark matter and the
charged mediators run in the loop.  In particular, in the limit where Standard Model
fermion helicity mixing is suppressed, the cross section for dark matter annihilation to various
final states is related to corrections to the Standard Model fermion charge form factor.  These corrections can be
extracted in a gauge-invariant manner from collider cross sections.
Although current measurements from colliders are not precise enough to provide useful constraints on
dark matter annihilation, improved measurements at future experiments, such as the International Linear Collider,
could improve these constraints by several orders of magnitude, allowing them to surpass the limits obtainable by
direct observation.
}
\maketitle

\section{Introduction}

It has long been appreciated that if dark matter (DM) is a Majorana fermion and
if the theory respects minimal flavor violation (MFV), then dark matter annihilation
to light Standard Model (SM) fermions is $p$-wave suppressed in the chiral limit.
Nevertheless, $p$-wave annihilation can be an important process in the early Universe during the epoch
of thermal dark matter freeze-out, when this velocity-suppression is only an ${\cal O}(0.1)$ effect.  In the current epoch,
when this velocity suppression is large ($\sim 10^{-6}$), indirect detection signals may instead by dominated by
other annihilation processes, such as virtual internal bremsstrahlung (IB), or annihilation to photons via a one-loop diagram.
In this work, we consider the constraints on these annihilation processes which can arise from associated corrections to the
vertex functions of the Standard Model fermions.

\begin{figure}[h]
\centering
\begin{tabular}{c}
\includegraphics[width=0.3\textwidth, keepaspectratio]{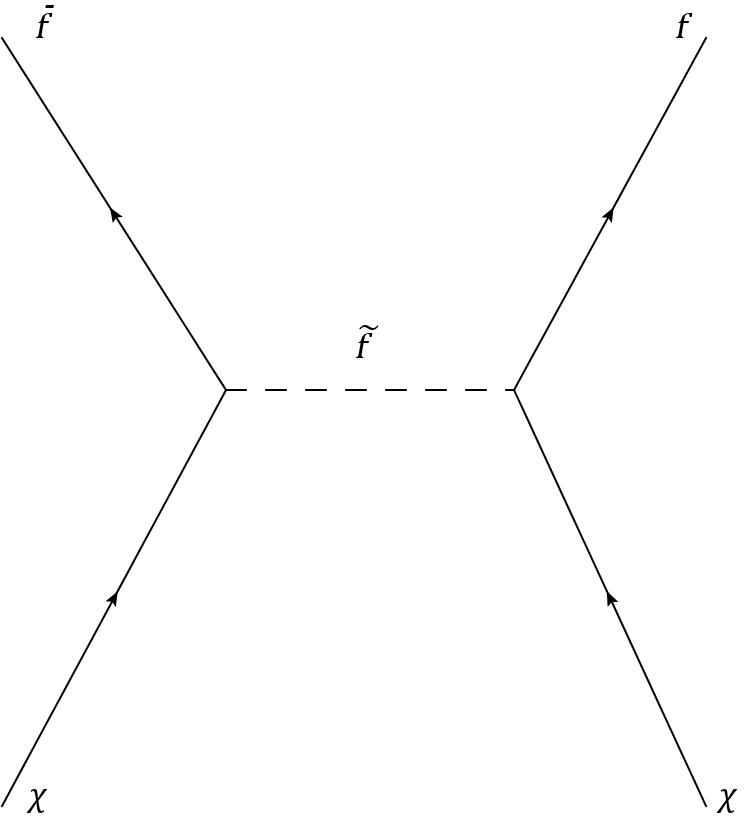}\hspace{2.0cm}
\includegraphics[width=0.3\textwidth, keepaspectratio]{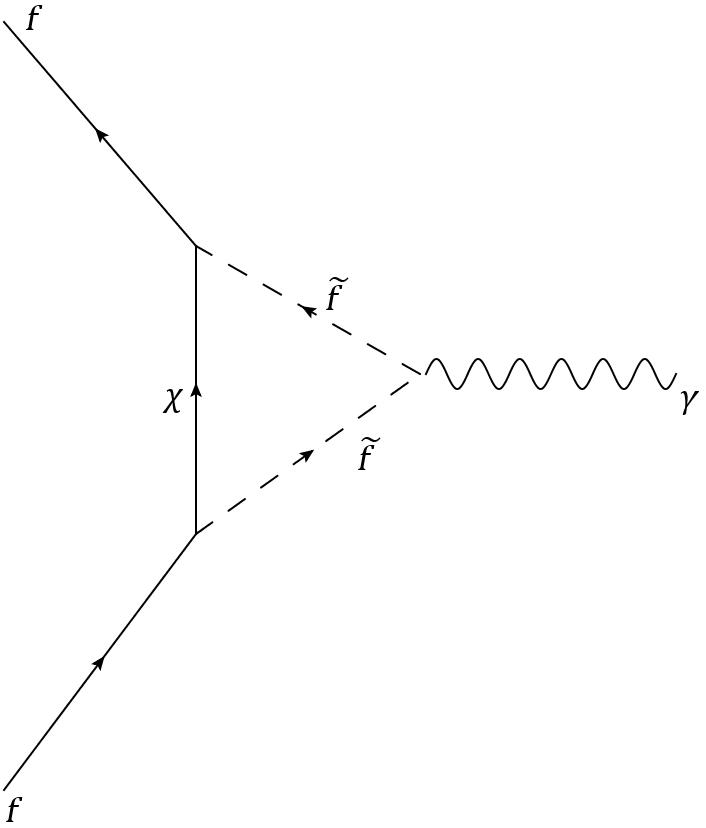}
\end{tabular}
\caption{ Feynman diagrams corresponding to the annihilation process $\chi \chi \rightarrow \bar f f$ (left diagram)
and to the correction to the fermion-photon vertex (right diagram).
\label{fig:FeynFFandChargeRadius} }
\end{figure}

If dark matter ($\chi$) annihilates to a charged SM fermion ($f$) through the $t$-channel exchange of charged mediating
particles, then there is necessarily a one-loop correction to the fermion-photon vertex arising
from diagrams where $\chi$ and the charged mediators run in the loop (see Figure~\ref{fig:FeynFFandChargeRadius}).
It has previously been shown~\cite{Fukushima:2013efa} that the $s$-wave dark matter annihilation matrix element
is directly correlated with the associated correction to the electric and magnetic dipole moments of the
SM fermions.  This connection follows from general principles: the annihilation process $\chi \chi \rightarrow \bar f f$
can only proceed from an $s$-wave initial state if there is some mixing of SM fermion helicities, and the
electromagnetic form factors which contribute to helicity mixing yield the electric and magnetic dipole moments
in the limit of small momentum transfer.  One might expect that the annihilation matrix elements
which do not mix SM fermion Weyl spinors can be similarly correlated with corrections to the electromagnetic form
factors which do not mix SM fermion Weyl spinors, and
we will see that this is indeed the case.
In particular, these annihilation matrix elements can be related to a correction to the helicity-preserving form factor,
$F_1 (q^2)$, in the limit of small momentum transfer.

The first moment of this form factor is sometimes parameterized by
a quantity referred to as the charge radius, and one might think of the one-loop correction involving dark matter
as providing a correction to the charge radius.  However, it is important to be precise about this concept; although
the form factor $F_1 (q^2)$ is gauge-invariant in quantum electrodynamics, it is generally not gauge-invariant in a
non-Abelian gauge theory (see, for example,~\cite{Lucio:1983mg,Monyonko:1984gb,Bernabeu:2004jr,Musolf:1990sa}),
such as the Standard Model.  Although the isolated diagram which we are interested in, involving dark
matter running in the loop, {\it is} gauge-invariant, there will be other corrections involving only SM particles which
are not.  But any $S$-matrix must be gauge-invariant, and we will find that one can extract constraints on the
correction in which we are interested from collider cross sections.

Constraints on corrections to the vertex functions can be inferred from cross section measurements at LEP~\cite{Acciarri:2000uh},
and future experiments may yield much more precise measurements.
We will find that collider constraints can be translated into bounds on the non-helicity-mixing terms of the cross sections for the dark matter annihilation
processes $\chi \chi \rightarrow \bar f f, \bar f f \gamma, \gamma \gamma$.
But the corrections to Standard Model cross sections at colliders involve all new physics, and can  arise both from diagrams
involving dark matter, and from new physics unrelated to dark matter.  Thus, it is always possible that
a large correction arising from loop corrections involving dark matter could be canceled by another large
correction arising from unrelated new physics, leaving a smaller total correction to a collider cross section.  As a result, the constraints
we will find, at a rigorous level, really indicate the extent to which one must fine-tune different corrections
in order to retain consistency with the data.

We will see that current bounds from LEP are not precise enough to be useful.
But future analyses, particularly using colliders with  higher energy and luminosity than LEP, could provide much better measurements of SM fermion
vertex functions, potentially improving the resulting constraints on dark matter
annihilation processes by several orders of magnitude.  In this case, the resulting constraints could not only be relevant for constraining
models of dark matter annihilation in the early Universe, but could exceed limits obtainable in the current epoch from direct observation.

The plan of this paper is as follows. In Section 2, we describe the relationship between the SM fermion vertex function correction and the
annihilation cross sections for the processes $\chi \chi \rightarrow \bar f f, \bar f f \gamma, \gamma \gamma $.  In section 3,
we determine the constraints imposed on dark matter annihilation, both
in the early Universe and in the present epoch, from current and potential measurements at colliders.
We conclude in section 4 with a discussion of our results.

\section{Connecting Dark Matter Annihilation to the Fermion Vertex Correction}

We first consider the scenario in which a Majorana fermion dark matter particle ($\chi$) annihilates to a
SM charged fermion/anti-fermion pair ($\bar f f$) through the $t$-channel exchange of mediating particles,
which necessarily are also charged.

The suppression of the $s$-wave annihilation cross section in the chiral limit ($m_f / m_\chi \rightarrow 0$) can
be understood from the conservation of angular momentum (see, for example,~\cite{Kumar:2013iva}).  Since the initial
state consists of two identical fermions, it must be totally anti-symmetric, implying that the $L=0$ initial state must
also have $S=0$, $J=0$.  The $\bar f f$ final state must then also have $J_z=0$, where we take the $z$-axis to be the direction
of motion of the outgoing particles.  This outgoing state necessarily has $L_z=0$, implying $S_z=0$.  So the outgoing
fermion and anti-fermion have the same helicity, and must arise from different Weyl spinors.  The $s$-wave annihilation matrix
element thus violates Standard Model flavor symmetries in the chiral limit, and must be proportional to the Weyl spinor mixing
introduced by new physics.  As a result, $s$-wave annihilation is suppressed in scenarios such as MFV, where Weyl spinor mixing vanishes
in the chiral limit.

The $L=1$ initial state, however, can have $S=1$, $J=1$; the $J=1$ fermion/anti-fermion final state
can have $S_z=J_z=1$, implying that the fermion and anti-fermion have opposite helicities and arise from the same Weyl spinor.  Thus,
the $p$-wave annihilation matrix element survives in the chiral limit even if there is no Weyl spinor mixing.

\begin{figure}[h]
\centering
\begin{tabular}{c}
\includegraphics[width=0.3\textwidth, keepaspectratio]{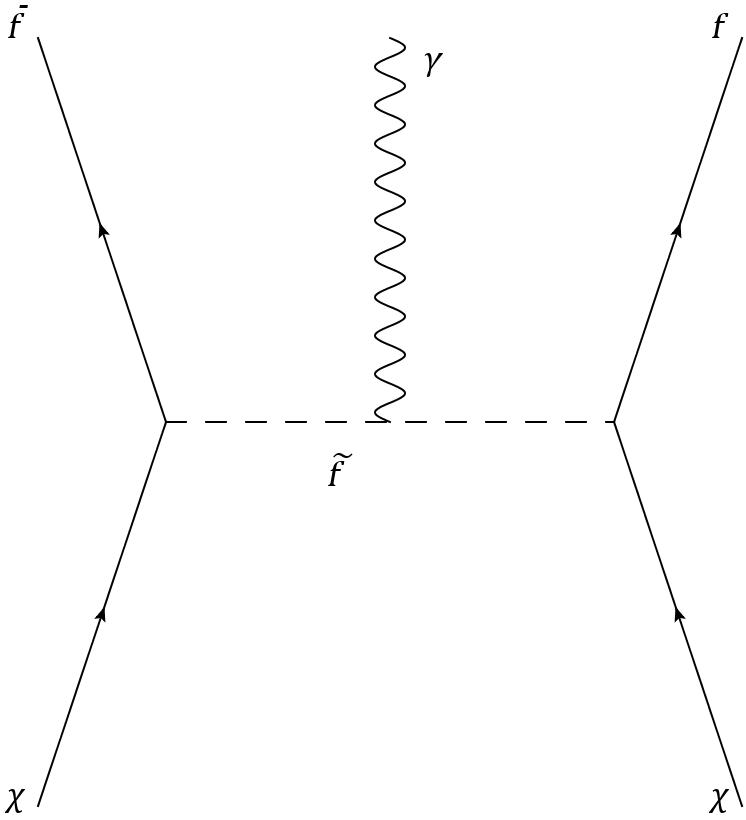}\hspace{2.0cm}
\includegraphics[width=0.3\textwidth, keepaspectratio]{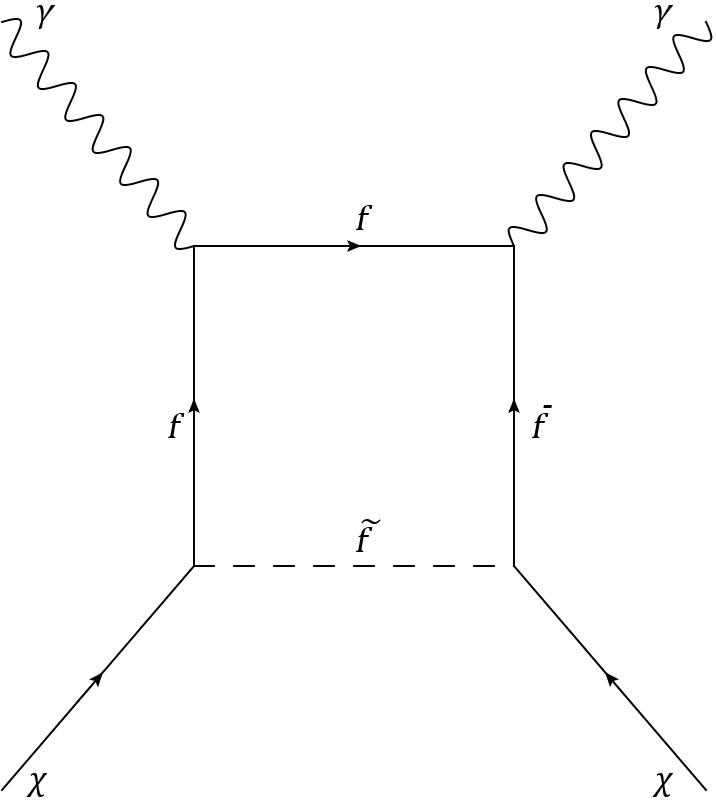}
\end{tabular}
\caption{ Representative Feynman diagrams corresponding to the annihilation processes $\chi \chi \rightarrow \bar f f \gamma$ (left diagram)
and $\chi \chi \rightarrow \gamma \gamma$ (right diagram).
\label{fig:FeynFFG and FeynGG} }
\end{figure}

However, the $s$-wave dark matter initial state can annihilate to a fermion/anti-fermion pair if an additional photon is also emitted.
In this case, the annihilation matrix element survives in the chiral limit; the $\bar f f \gamma$ final state can have $J=0$ even if $f$
and $\bar f$ arise from the same Weyl spinor, due to the angular momentum carried by the photon.  Similarly, the $s$-wave dark matter
initial state can annihilate to a $\gamma \gamma$ final state through a one-loop diagram with $f$ and charged mediators running in the
loop; this amplitude also survives in the chiral limit (see Figure~\ref{fig:FeynFFG and FeynGG}).

In general, we can express the fermion-photon vertex as
\bea
\Gamma^\mu (q^2) &=& \gamma^\mu F_1 (q^2) + \frac{\imath \sigma^{\mu \nu} q_\nu }{2m} F_2 (q^2)
+ \frac{\imath \sigma^{\mu \nu} \gamma^5 q_\nu  }{2m} F_3 (q^2)
+ (\gamma^\mu q^2 - \Dsl q q^\mu) \gamma^5 F_A (q^2) ,
\eea
where $q$ is the photon momentum and the $F_{1,2,3,A}$ are form factors.  The form factors can
receive contributions from new physics, including one-loop vertex correction diagrams where $\chi$ and the charged mediators run
in the loop.  One can classify these form factors based on the transformation properties of the various vertex correction terms under $C$, $P$, and SM
flavor symmetries.  In particular, $F_2$ and $F_3$ violate SM flavor symmetries by mixing SM Weyl spinors; they must vanish in
the chiral limit if there is no Weyl spinor mixing.  Both $F_1$ and $F_2$ are invariant under $C$ and $P$ separately, while $F_3$ is
$CP$-odd.  $F_A$ is $CP$-even, but is odd under $C$ and $P$ individually.

We can thus make a general connection between corrections to the various form factors and terms in the $\chi \chi \rightarrow \bar f f,
\bar f f \gamma, \gamma \gamma$
annihilation matrix elements in the chiral limit.  The non-relativistic $s$-wave annihilation matrix element for the process
$\chi \chi \rightarrow \bar f f$ is necessarily proportional to Weyl spinor
mixing, so these terms will be related to corrections to the $F_2$ and $F_3$ form factors, and specifically to corrections to the magnetic
and electric dipole moments~\cite{Fukushima:2013efa}.
In particular, the $CP$-conserving term of the $s$-wave annihilation matrix element is related to the correction to
magnetic dipole moment, while the $CP$-violating term is related to the electric dipole moment correction.  But these terms vanish in the
chiral limit if there is no Weyl spinor mixing.

If there is no Weyl spinor mixing, and if $C$ and $P$ are separately conserved, then the only form
factor which can receive a correction is $F_1$.  This correction is then related to the annihilation matrix elements
which do not mix SM fermion Weyl spinors, in particular, the $p$-wave matrix element for the process $\chi \chi \rightarrow \bar f f$
and the $s$-wave matrix elements for the processes $\chi \chi \rightarrow \bar f f \gamma, \gamma \gamma$.\footnote{Note, we do not
consider the $p$-wave matrix elements for the processes $\chi \chi \rightarrow \bar f f \gamma, \gamma \gamma$, since they are always
suppressed, compared to their $s$-wave counterparts.}

We can implement the scenario described above with a simplified model, in which $\chi$ is a gauge-singlet
Majorana fermion which couples to a single Standard Model fermion $f$ through the
exchange of two scalars, $\tilde f_L$ and $\tilde f_R$, via the Lagrangian
\bea
{\cal L}_f &=& \lambda_f \tilde f_L (\bar \chi P_L f) + \lambda_f \tilde f_R (\bar \chi P_R f) + h.c.,
\label{eq:Lagrangian}
\eea
where $\lambda_f$ is a real dimensionless coupling constant. We assume that $\chi$ and $\tilde f_{L,R}$ are odd
under a $Z_2$ symmetry which stabilizes the dark matter.
The $\tilde f_{L,R}$ have the same quantum numbers as left- and right-handed sfermions.
We further assume that $\tilde f_L$ and $\tilde f_R$ do not
mix and have the same mass ($m_{\tilde f} \equiv m_{\tilde f_L} = m_{\tilde f_R}$).  These choices ensure that
the annihilation matrix elements respect $C$ and $P$, and do not mix SM Weyl spinors in the chiral limit.

Note, for simplicity, we have assumed that the dominant dark matter coupling is to a single Standard Model fermion.
Of course, since $\tilde f_L$ is a member of an $SU(2)_L$ doublet, it has a corresponding partner, with a different
charge, which we may denote by $\tilde f'_L$.  One could embed the coupling of this field to dark matter within
a $C$- and $P$-invariant theory by adding a copy of the interaction Lagrangian in eq.~(\ref{eq:Lagrangian}),
\bea
{\cal L}_{f'} &=& \lambda_{f'} \tilde f'_L (\bar \chi P_L f') + \lambda_{f'} \tilde f'_R (\bar \chi P_R f') + h.c.,
\label{eq:Lagrangian_SU(2)}
\eea
where $f'$ is a fermion in the same generation as $f$, $\tilde f'_R$ is another scalar field,
and $\lambda_{f'}$ is a dimensionless coupling.
For most of the quantities which we will consider there will be no interference between the terms in
${\cal L}_f$ and ${\cal L}_{f'}$.
In particular,
corrections to the fermion-photon vertex for $f$, and the cross sections for the processes $\chi \chi \rightarrow \bar f f$
and $\chi \chi \rightarrow \bar f f \gamma$, are unchanged, while similar processes involving $f'$ depend only on
the interactions contained in ${\cal L}_{f'}$.  But the cross section for the
process $\chi \chi \rightarrow \gamma \gamma$ will depend on all of these terms, since $f$ and $f'$ can both
run in the loop; note, though, even this consideration is not relevant if $f$ is a charged lepton,
since $f'$ and $\tilde f'_{L,R}$ are then electrically neutral and do not couple to the photon.

One could also consider the case where $\tilde f_{L,R}$ couples to multiple fermions in different generations,
but in this case, there can also be contributions to tightly constrained flavor-changing processes,
such as $f_1 \rightarrow f_2 \gamma$, through diagrams where $\tilde f_{L,R}$ and $\chi$ run in the loop.
As a result, one might expect only the coupling of $\tilde f_{L,R}$ to one of the fermions to be dominant.
We will discuss
the effect on this analysis if dark matter couples to multiple fermions in the next section.

\subsection{Annihilation}

To lowest order in the relative velocity ($v$), the annihilation cross section for the process $\chi \chi \rightarrow \bar f f$ can
be written as (see, for example,~\cite{Bell:2011if,Pierce:2013rda,Fukushima:2014yia})
\bea
\langle \sigma_{\bar f f} v \rangle &=& {\lambda_f^4 N_c r \over 12\pi m_{\tilde f}^2} \frac{ 1+r^2 }{(1+r)^4} \langle v^2 \rangle ,
\label{eq:ff_crosssection}
\eea
where $\langle v^2 \rangle$ is the thermal average of $v^2$ and $r \equiv  m_\chi^2 / m_{\tilde f}^2$ satisfies
the constraint $0 < r \leq 1$.  Since there is no Weyl spinor mixing, this cross section is necessarily $p$-wave suppressed.
$N_c$ is the number of color states for the fermion $f$; $N_c=1$ if $f=e,\mu,\tau$, while $N_c=3$ if $f=q$.
If the dark matter is a thermal relic, then at the time of freeze-out, one expects
$\langle v^2 \rangle \sim 0.1$.  But in the current epoch, $\langle v^2 \rangle \sim 10^{-6}$, implying that the
$\bar f f \gamma$ and $\gamma \gamma$ final states may be important for the purposes of indirect detection.

The differential cross section for the process $\chi \chi \rightarrow \bar f f \gamma$ can be written as~\cite{Bringmann:2007nk,Kumar:2016cum}
\bea
\frac{d\sigma_{\bar f f \gamma} }{dx} v  &=&\frac{\lambda_f^4 N_c \alpha Q_f^2 r (1-x)}{32 \pi^2 m_{\tilde f}^2}
\left[\frac{4x}{(1+r)(1+r-2xr)} -\frac{2x}{(1+r-xr)^2}
\right.
\nonumber\\
&\,& \left.
+\frac{(1+r)(1+r-2xr)}{r(1+r-xr)^3} \ln \frac{1+r-2xr}{1+r} \right] ,
\eea
where $Q_f$ is the electric charge of $f$, $x \equiv E_\gamma / m_\chi$,
$E_\gamma$ is the energy of the photon, and $0 \leq x \leq 1$.
The total cross section is then given by~\cite{Bell:2011if,Fukushima:2012sp,Kumar:2016cum}
\bea
\langle \sigma_{\bar f f \gamma} v \rangle &=& \frac{\lambda_f^4 N_c \alpha Q_f^2}{32 \pi^2 m_{\tilde f}^2 r^2}
\left[(1+r)\left(\frac{\pi^2}{6} - \ln^2 \frac{1+r}{2} - 2\mathrm{Li}_2 \frac{1+r}{2} \right)
\right.
\nonumber\\
&\,& \left.
+ \frac{3r^2 + 4r}{1+r} +\frac{4-3r-r^2}{2} \ln \frac{1-r}{1+r} \right] .
\label{eq:ffgamma_crosssection}
\eea
We are primarily interested in the limits $r \sim 1$ (when the $\chi$ has nearly the same mass as the
mediators) and $r \ll 1$ (when the mediators are heavy).  We then find
\bea
\langle \sigma_{\bar f f \gamma} v \rangle_{r \ll 1} &\sim& \frac{\lambda_f^4 N_c r }{120 \pi^2 m_{\tilde f}^2 } (\alpha Q_f^2 r^2) ,
\nonumber\\
\langle \sigma_{\bar f f \gamma} v \rangle_{r \rightarrow 1} &\sim& \frac{\lambda_f^4 N_c }{32 \pi^2 m_{\tilde f}^2 }
\left[\frac{7}{2} -\frac{\pi^2}{3} \right] \alpha Q_f^2  .
\eea
Note that, even in the $r \rightarrow 1$ limit, $\langle \sigma_{\bar f f \gamma} v \rangle$ is
suppressed relative to $\langle \sigma_{\bar f f} v \rangle$ by a factor
$(3/\pi)[7/2 - \pi^2/3] (\alpha / \langle v^2 \rangle ) Q_f^2 $.  Thus, the process
$\chi \chi \rightarrow \bar f f \gamma$ will necessarily be subleading at the time of
freeze-out, when $\alpha / \langle v^2 \rangle \sim {\cal O}(0.1)$.

The total cross section for the one-loop process $\chi \chi \rightarrow \gamma \gamma$ can be written as
\cite{Bergstrom:1997fh,Bern:1997ng,Ullio:1997ke,Kumar:2016cum}
\bea
\langle \sigma_{\gamma \gamma} v \rangle &=& \frac{\lambda_f^4 N_c^2 \alpha^2 Q_f^4 }{64 \pi^3 m_{\tilde f}^2 r}
\left[\mathrm{Li}_2 (r) - \mathrm{Li}_2 (-r) \right]^2 ,
\label{eq:gammagamma_crosssection}
\eea
yielding
\bea
\langle \sigma_{\gamma \gamma} v \rangle_{r \ll 1} &\sim& \frac{\lambda_f^4 N_c^2 r  }{16 \pi^3 m_{\tilde f}^2 } \alpha^2 Q_f^4 ,
\nonumber\\
\langle \sigma_{\gamma \gamma} v \rangle_{r \rightarrow 1} &\sim& \frac{\lambda_f^4 N_c^2 \pi  }{1024 m_{\tilde f}^2 } \alpha^2 Q_f^4  .
\eea

\subsection{Correction to the Form Factor}

One can also compute the vertex correction arising from the two diagrams where $\chi$ and either $\tilde f_L$ or $\tilde f_R$ run in
the loop.  As expected, the only form factor which receives a correction is $F_1$.
To first order in $q^2$, $F_1$ is often represented by the expansion
\bea
F_1 (q^2) &\equiv& 1 + \frac{1}{6} q^2 R^2 +{\cal O}(q^4),
\eea
where $R$ is referred to as the charge radius (gauge-invariance implies that $F_1 (q^2=0) =1$).
In a non-Abelian gauge theory, however, this quantity is not necessarily gauge-invariant
for $q^2 \neq 0$~\cite{Lucio:1983mg,Monyonko:1984gb,Bernabeu:2004jr,Musolf:1990sa},
leading to ambiguity in the meaning of the charge radius.
For example, if one computed in an $R_\xi$-gauge,
one would find that, although the diagram in the right panel of Figure~\ref{fig:FeynFFandChargeRadius} is
$\xi$-independent (as it contains no gauge bosons in the loop), there are other vertex corrections involving
Standard Model fields which are $\xi$-dependent; the $\xi$-dependence cancels when computing an observable
quantity, such as an $S$-matrix element.

But the vertex correction arising from the diagram in Figure~\ref{fig:FeynFFandChargeRadius} (right panel) yields a correction
to $F_1$ which is gauge-invariant, and can be parameterized as
\bea
\Delta F_1 (q^2) &=& \frac{1}{6} q^2 \Delta R^2 + {\cal O}(q^4) .
\eea
The ${\cal O}(q^0)$ term in the correction is necessarily canceled by
the fermion external leg correction, as guaranteed by the Ward-Takahashi identity.
The parameter $\Delta R$ may be thought of heuristically as a correction to the charge radius, but it is
more precisely a parameter which can be extracted from cross sections.

This correction is give by
\bea
|\Delta R^2 | &=& \frac{3\lambda_f^2}{8\pi^2 m_{\tilde f}^2} \frac{3r^2+(1-4r)-2r^2 \ln r}{4(1-r)^3} .
\eea
Again, our main interest is in the limits $r \sim 1$ and $r \ll 1$, in which case we find
\bea
\left| \Delta R^2 \right|_{r \ll 1} &=& \frac{3\lambda_f^2}{32\pi^2 m_{\tilde f}^2} ,
\nonumber\\
\left| \Delta R^2 \right|_{r \sim 1} &=& \frac{\lambda_f^2}{16\pi^2 m_{\tilde f}^2} .
\eea
We thus see that the correction scales
\bea
\Delta R^2 \propto \lambda_f^2 / m_{\tilde f}^2 ,
\eea
where the remaining model dependence arises only from rescaling by an ${\cal O}(1)$ function of $r$.

We may then express the relevant annihilation cross sections in terms of $\Delta R^2$ and $r$ as
\bea
\langle \sigma_{\bar f f} v \rangle &=& \left[N_c \left(\Delta R^2 \right)^2 m_{\tilde f}^2 r \langle v^2 \rangle \right]
\left(\frac{256\pi^3}{27}\left(\frac{(1-r)^3}{3r^2+(1-4r)-2r^2 \ln r} \right)^2 \left(\frac{ 1+r^2}{(1+r)^4} \right) \right)    ,
\nonumber\\
\langle \sigma_{\bar f f \gamma} v \rangle &=& \left[N_c \left(\Delta R^2 \right)^2 m_{\tilde f}^2 r^3 \alpha Q_f^2   \right]
\left(\frac{32\pi^2}{9}\left(\frac{(1-r)^3}{3r^2+(1-4r)-2r^2 \ln r} \right)^2 \left( \frac{1}{r^5}\right) \right)
\nonumber\\
&\,&  \times
\left[(1+r)\left(\frac{\pi^2}{6} - \ln^2 \frac{1+r}{2} - 2\mathrm{Li}_2 \frac{1+r}{2} \right)
+ \frac{3r^2 + 4r}{1+r} +\frac{4-3r-r^2}{2} \ln \frac{1-r}{1+r} \right] ,
\nonumber\\
\langle \sigma_{\gamma \gamma} v \rangle &=& \left[N_c^2 \left(\Delta R^2 \right)^2 m_{\tilde f}^2 r \alpha^2 Q_f^4  \right]
\left(\frac{16\pi}{9}\left(\frac{(1-r)^3}{3r^2+(1-4r)-2r^2 \ln r} \right)^2 \right)
\left[\frac{\mathrm{Li}_2 (r) - \mathrm{Li}_2 (-r)}{r} \right]^2  .
\nonumber\\
\label{eq:Constraint}
\eea
All of these annihilation cross sections scale with $\Delta R^2$ and the mediator mass as
$\langle \sigma v \rangle \propto (\Delta R^2)^2 m_{\tilde f}^2$; constraints on the annihilation
cross section thus become weaker if the mediating particles are heavy.  But the cross sections for the
processes $\chi \chi \rightarrow \bar f f, \bar f f \gamma, \gamma \gamma$ scale as $r$, $r^3$, $r$,
respectively.  These constraints are thus most stringent when the dark matter is much lighter than the
mediator, and the mediator is as light as is consistent with experimental constraints.

Note, we have assumed that dark matter couples to only one SM fermion, so there are no interference effects.
If dark matter
couples to multiple SM fermions $f_i$, then the connection between $\Delta R^2$ for the fermion
$f_i$ and the cross sections $\langle \sigma_{\bar f_i f_i} v \rangle$ and $\langle \sigma_{\bar f_i f_i \gamma} v \rangle$
is unchanged, since in each case only $f_i$ appears in the relevant diagrams.
However, for the loop diagrams which contribute to the process
$\chi \chi \rightarrow \gamma \gamma$, all of the $f$s can run in the loop.

We can consider the simple generalization in which, each SM fermion $f_i$ couples to $\chi$
via an interaction with the scalars $\tilde f_{Li, Ri}$, with coupling $\lambda _{f_i}$.
Assuming  that
$m_{f_i} \ll m_{\chi}, m_{\tilde f_i}$ for all $i$, then we find (generalizing the result from~\cite{Kumar:2016cum})
\bea
\langle \sigma_{\gamma \gamma} v \rangle &=& \frac{\alpha^2}{64 \pi^3 m_\chi^2}
\left(\sum_i \lambda_{f_i}^2 N_{C(i)} Q_{f_i}^2 \left[\mathrm{Li}_2 (r_i) - \mathrm{Li}_2 (-r_i) \right] \right)^2 ,
\eea
where $r_i = m_\chi^2 / m_{\tilde f_i}^2$.

\section{Constraints from Data}

We have found relationships between dark matter annihilation cross sections and the vertex correction diagram
given in Figure~\ref{fig:FeynFFandChargeRadius} (right panel), but in order to constrain the vertex correction, it
is necessary to relate it to observables.  This is non-trivial because in a non-Abelian gauge
theory, such as the Standard Model, $F_1 (q^2)$ need not be gauge-invariant if
$q^2 \neq 0$~\cite{Lucio:1983mg,Monyonko:1984gb,Bernabeu:2004jr,Musolf:1990sa}.  However, the gauge-dependence cancels
when computing an observable
quantity, such as an $S$-matrix element.  We will see that the correction given by the diagram in the right panel
of Figure~\ref{fig:FeynFFandChargeRadius}, and in particular, the parameter $\Delta R^2$, can be extracted from observable cross sections.

For simplicity and concreteness, we will focus on the case where $f=\mu$.
The vertex correction can be extracted from two cross sections: the cross sections for the process
$e^+ e^- \rightarrow \bar \mu \mu$ and for the process $e^+ e^- \rightarrow \bar \tau \tau$, for example.  At tree-level, both processes are mediated by
neutral gauge boson ($\gamma, Z$) exchange in the $s$-channel (we may neglect Higgs boson exchange, as the coupling of the
Higgs boson to the electron is very small).  If we take the limit $m_{\tau, \mu} \ll s$, then the Standard Model cross section
is flavor independent:
\bea
\sigma_{e^+ e^- \rightarrow \bar \mu \mu}^{SM}  &=& \sigma_{e^+ e^- \rightarrow \bar \tau \tau}^{SM} .
\eea
At one-loop level, these cross sections will involve a variety of Standard Model corrections to the
fermion vertex functions, including gauge-dependent corrections to the $F_1$ and $F_A$ form factors,
though the gauge-dependence will cancel in the full cross sections.  But the point is that these
corrections will be the same for both cross sections, and will cancel when the difference is taken.

If we include the new fields and interactions encoded in ${\cal L}_{f=\mu}$, then we find two new classes
of one-loop diagrams which can contribute to these processes.  These are diagrams where a neutral gauge boson is exchanged
in the $s$-channel and either a) the gauge boson two-point function is corrected by $\tilde f_{L,R}$ running in a loop, or
b) the muon vertex or propagator is corrected by a $\tilde f_{L,R}$ and $\chi$ running in the loop.  The first set of diagrams provide
the same correction to the matrix element for $e^+ e^- \rightarrow \bar \mu \mu$ and $e^+ e^- \rightarrow \bar \tau \tau$.
But the vertex and fermion propagator correction diagrams affect only the cross section for $e^+ e^- \rightarrow \bar \mu \mu$, so
the correction to the vertex can be extracted from a search for a deviation from lepton universality.

The effect of the muon vertex and propagator corrections on the annihilation matrix element can be parameterized by a 
$q^2$-dependent rescaling of the $e^+ e^- \rightarrow \bar \mu \mu$ matrix element by the factor:
\bea
1 + \frac{1}{6} q^2 \Delta R^2 + {\cal O}(q^4 \Delta R^4)^2 ,
\eea
where $q^2=s$.
The Ward-Takahashi identity ensures that, after including both the vertex and propagator correction, the
rescaling becomes trivial at $q^2 = 0$.
Note that both the $\gamma\mu\mu$ and $Z\mu \mu$ vertices are corrected.  But since we have introduced no
chirality mixing, and the new physics contributions respect $P$, the corrections to the $\mu_L$ and $\mu_R$
vertices must be the same, so the $Z$-vertex is rescaled by the same factor as the $\gamma$-vertex.

If we assume that the one loop corrections are small, then the vertex correction diagram only affects the cross section
through interference with the tree-level matrix element.  We thus find
\bea
\sigma_{e^+ e^- \rightarrow \bar \mu \mu}^{1-loop} - \sigma_{e^+ e^- \rightarrow \bar \tau \tau}^{1-loop}
&\approx& \left( \sigma_{e^+ e^- \rightarrow \bar \mu \mu}^{SM(tree)} \right) \times (q^2 \Delta R^2 /3).
\eea
The left hand side of this relation involves cross sections calculated at one-loop, including the
effects of new physics, and can be measured or constrained using collider data.  The right hand side is
a function of the parameter $\Delta R^2$ and of tree-level SM cross sections which can be easily computed,
allowing one to extract $\Delta R^2$ from data.

Although we focus specifically on the case where $f=\mu$, and on extracting the parameter $\Delta R^2$
from the processes $e^+ e^- \rightarrow \bar \mu \mu, \bar \tau \tau$, this strategy can clearly be
generalized to other choices of $f$, and to other initial states, such as $e^- q$.
We will comment in the Conclusions
on more complicated scenarios, in which extracting the correction unambiguously may not be as easy.

\subsection{Collider Results}

Searches for sleptons have been performed at LEP~\cite{Heister:2002jca,Achard:2003ge,Abdallah:2003xe,Abbiendi:2003ji}
and the LHC~\cite{Aad:2014vma,Khachatryan:2014qwa}, and can be used to place rough
lower bounds on $m_{\tilde f}$ in the case where $f=e, \mu, \tau$.  Consistency with searches at LEP require $m_{\tilde f} \gtrsim 100\gev$,
while LHC searches would require  $m_{\tilde f} \gtrsim 300\gev$ in the $r \rightarrow 0$ limit (these
constraints are a little weaker in the $r \rightarrow 1$ limit).
In the case where $f=q$, one can estimate lower bounds on $m_{\tilde f}$ from LHC squark searches~\cite{Aad:2014wea}.
These bounds would roughly require $m_{\tilde f} \gtrsim 500\gev$ in the $r \rightarrow 0$ limit (again, these
constraints are somewhat weaker in the $r \rightarrow 1$ limit).
As shown in eq.~(\ref{eq:Constraint}), bounds on the dark matter annihilation cross sections
are strongest when $m_{\tilde f}$ saturates the lower bound.

LEP has made precise measurements of the energy dependence of the $e^+ e^- \rightarrow \bar f f$
cross section~\cite{Acciarri:2000uh}.
For the LEP analysis, $f=e, \mu, \tau$, or $q$, where $q=u,d,s,c,b$ (it was assumed that quarks have
identical vertex functions).
The correction to the cross section was parameterized by
\bea
\frac{d\sigma_{e^+ e^- \rightarrow \bar f f} }{dq^2} &=& \frac{d\sigma_{e^+ e^- \rightarrow \bar f f}^{SM} }{dq^2}
\left(1+\frac{q^2 \Delta R^2}{6} \right)^4 .
\eea
These constraints on $|\Delta R|$ are summarized in Table~\ref{Tab:ChargeRadiiLimits}.

\begin{table}[h]
\centering
\begin{tabular}{|c|c|}
  \hline
  channel & $ |\Delta R|$ \\
  \hline
  $e^+ e^-$ & $<3.1 \times 10^{-19}$m \\
  $\mu^+ \mu^-$ & $<2.4 \times 10^{-19}$m \\
  $\tau^+ \tau^-$ & $<4.0 \times 10^{-19}$m \\
  $\bar q q$ & $<3.0 \times 10^{-19}$m \\
  \hline
\end{tabular}
\caption{\label{Tab:ChargeRadiiLimits} Constraints on the parameter $|\Delta R|$ for Standard Model fermions ($e$, $\mu$, $\tau$)
and quarks ($q$) obtained from searches at LEP~\cite{Acciarri:2000uh}, using the parametrization discussed in the text.}
\end{table}

Note, it was assumed that the electron and outgoing fermion vertex functions received the same correction; these
constraints are thus weakened for $f \neq e$ by a factor $\sqrt{2}$ if the electron is instead taken to be point-like.
Comparable constraints on the correction to the quark vertex functions have been obtained at HERA~\cite{Abramowicz:2016xzf},
using a similar parametrization.

If $m_{\tilde f} \gtrsim 300\gev$, then a value of $\Delta R^2$ large enough to be constrained by LEP data would
already be close to (if not ruled out by) the perturbativity limit, $\lambda_f \leq \sqrt{4\pi}$.  It is thus already clear that,
given the constraints on charged mediators from the LHC, current bounds from LEP can only be of limited utility in
constraining dark matter annihilation.

The parametrization used by LEP  for the correction to the cross section is different from the one
which we have used, implying that their constraints cannot directly be translated into bounds on dark matter annihilation.
However, LEP bounds the deviation of the cross section from the Standard Model prediction;
given that their bounds on the deviations of the $e^+ e^- \rightarrow \bar \mu \mu$ and $\rightarrow  \bar \tau \tau$
cross sections are similar to each other,
the deviation of these cross sections from each other cannot be much larger than the deviation from the SM prediction.
The bound on the parameter $\Delta R^2$ for the muon, as we have parametrized it, thus cannot be much larger than
that found by LEP using their parametrization.  In any case, this
bound is already too weak to usefully constrain dark matter annihilation.
But future analyses at experiments with higher energy and luminosity could yield
much more precise measurements of these cross sections.

Note that, since the $Z \mu \mu$ vertex is also corrected, one can also constrain $\Delta R$ through measurements of the
lepton universality of $Z$-boson decays.  For $\Delta R \sim {\cal O}(10^{-19}~\m)$, the deviation from lepton universality
in $Z$-decay branching fractions would be ${\cal O}(10^{-3})$, which is comparable to the current bound from LEP data at the
$Z$-pole~\cite{ALEPH:2005ab}.

For the remainder of this section, we continue to focus on the case
$f=\mu$, $m_{\tilde f}=300\gev$ and consider two different scenarios: bounds on $\Delta R$ for the muon at the level given by LEP,
and the possibility that the bound on $\Delta R$ could be improved by a factor of 10 with future experiments.
Since the annihilation cross section bounds arising from collider measurements scale as
$\langle \sigma_{\bar f f, \bar f f \gamma, \gamma \gamma} v \rangle \propto (\Delta R)^4 m_{\tilde f}^2$,
one can easily rescale these bounds for a different choice of $m_{\tilde f}$, or for future constraints on $\Delta R$.
Similarly, the maximum cross section consistent with perturbativity (for any of the channels we have discussed)
scales as $\propto m_{\tilde f}^{-2}$; one can easily rescale the perturbativity limit for a different choice of
$m_{\tilde f}$.

\subsection{Thermal Freeze-out}

At the time of dark matter thermal freeze-out, one typically expects~\cite{Freezeout} $\langle v^2 \rangle \sim {\cal O}(0.1)$.
We can express the $\chi \chi \rightarrow \bar f f$ $p$-wave annihilation cross section as
\bea
\langle \sigma_{\bar f f} v \rangle &=&
(5.5 \times 10^3~\text{pb} ) r N_c
\left[\left(\frac{\Delta R^2}{(3 \times 10^{-19}~\m)^2} \right)^2 \left(\frac{m_{\tilde f}}{300\gev}\right)^2  \frac{\langle v^2 \rangle}{0.1} \right]
\nonumber\\
&\,& \times
\left(\frac{(1-r)^3}{3r^2+(1-4r)-2r^2 \ln r} \right)^2 \left(\frac{ (1+r^2)}{(1+r)^4} \right)    .
\eea

\begin{figure}[ht]
\centering
\includegraphics[width=0.8\textwidth, keepaspectratio]{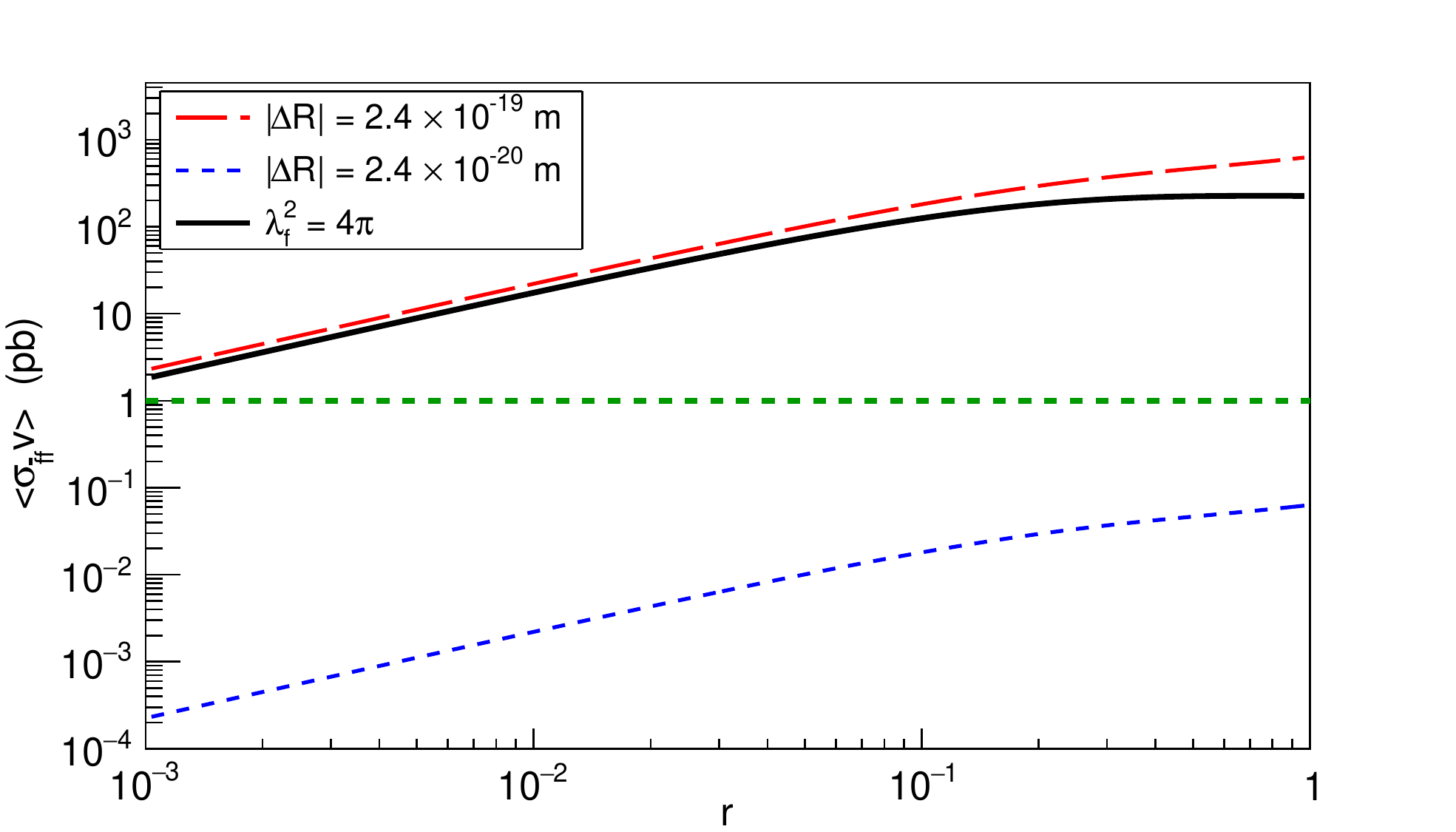}\hspace{0.5cm}
\caption{\label{fig:ff} Bounds on $\langle \sigma_{\bar f f} v \rangle$ (in pb) arising
from measurements of $\Delta R$, where $f=\mu$,
$m_{\tilde f}=300\gev$ and $\langle v^2 \rangle =0.1$.  The red long-dashed contour corresponds to the case
where $\Delta R = 2.4 \times 10^{-19}~\m$ (the current limit reported by LEP~\cite{Acciarri:2000uh}), and the
blue short-dashed contour corresponds to the case where $\Delta R = 2.4 \times 10^{-20}~\m$.
The region above the curves
is excluded, if there are no other corrections arising from new physics unconnected to the model we have discussed.  Also shown is the
perturbativity limit, $\lambda_f^2=4\pi$ (black solid).  The green short-dashed line at $1~\pb$ represents
the approximate annihilation cross section required to achieve the standard thermal abundance.}
\end{figure}

In Figure~\ref{fig:ff}, we plot the constraints on $\langle \sigma_{\bar f f} v \rangle$ as a function of $r$ ($m_{\tilde f}=300\gev,
\langle v^2 \rangle =0.1$).
The red long-dashed contour describes the constraints which arise from LEP bounds on $\Delta R$ for the muon,
while the blue short-dashed contour describes the constraints which would
arise if $\Delta R$ were constrained to be 10 times smaller by future data.
The regions above the contours are excluded, assuming
that there is no other source for vertex corrections.
Also plotted are the largest cross section consistent with perturbativity ($\lambda_f^2 =4\pi$),
and the approximate thermal abundance cross section ($1~\pb$).
We see that, if the precision of constraints on $\Delta R$ can reach the ${\cal O}(10^{-20}~\m)$ level for $m_{\tilde f} \lesssim 3000\gev $,
then models in which the dark matter relic density in the
early Universe is depleted by two-body $p$-wave annihilation would be correlated with a measurable
energy-dependent deviation from lepton universality.

\subsection{Indirect Detection in the Current Epoch}

In the current epoch, when $\langle v^2 \rangle \sim 10^{-6}$, indirect detection signals arising from
the processes $\chi \chi \rightarrow \bar f f \gamma$  (internal bremsstrahlung) and $\chi \chi \rightarrow
\gamma \gamma$ (via a one-loop diagram) may be competitive with $\chi \chi \rightarrow \bar f f$.
We may express these annihilation cross sections as
\bea
\langle \sigma_{\bar f f \gamma} v \rangle &=& (12.8~\text{pb})r^3 N_c Q_f^2
\left[ \left(\frac{\Delta R^2}{(3 \times 10^{-19}~\m)^2} \right)^2
\left(\frac{m_{\tilde f}}{300\gev}\right)^2    \right]
\nonumber\\
&\,& \times
\left(\frac{15}{4}\left(\frac{(1-r)^3}{3r^2+(1-4r)-2r^2 \ln r} \right)^2 \left( \frac{1}{r^5}\right) \right)
\nonumber\\
&\,&  \times
\left[(1+r)\left(\frac{\pi^2}{6} - \ln^2 \frac{1+r}{2} - 2\mathrm{Li}_2 \frac{1+r}{2} \right)
+ \frac{3r^2 + 4r}{1+r} +\frac{4-3r-r^2}{2} \ln \frac{1-r}{1+r} \right] ,
\nonumber\\
\langle \sigma_{\bar \gamma \gamma} v \rangle &=& (0.22~\text{pb})r N_c^2 Q_f^4
\left[\left(\frac{\Delta R^2}{(3 \times 10^{-19}~\m)^2} \right)^2
\left(\frac{m_{\tilde f}}{300\gev}\right)^2   \right]
\nonumber\\
&\,& \times
\left(\frac{(1-r)^3}{3r^2+(1-4r)-2r^2 \ln r} \right)^2
\left(\frac{\mathrm{Li}_2 (r) - \mathrm{Li}_2 (-r)}{2r} \right)^2  .
\eea
Note, even though we might expect that $\chi$  couples to $\nu_\mu$ through interactions
with the other member of the $SU(2)_L$ doublet of which $\tilde f_L$ is a member, those
interactions will not contribute to $\chi \chi \rightarrow \gamma \gamma$, since
neutrinos are electrically neutral.

\begin{figure}[ht]
\centering
\includegraphics[width=0.8\textwidth, keepaspectratio]{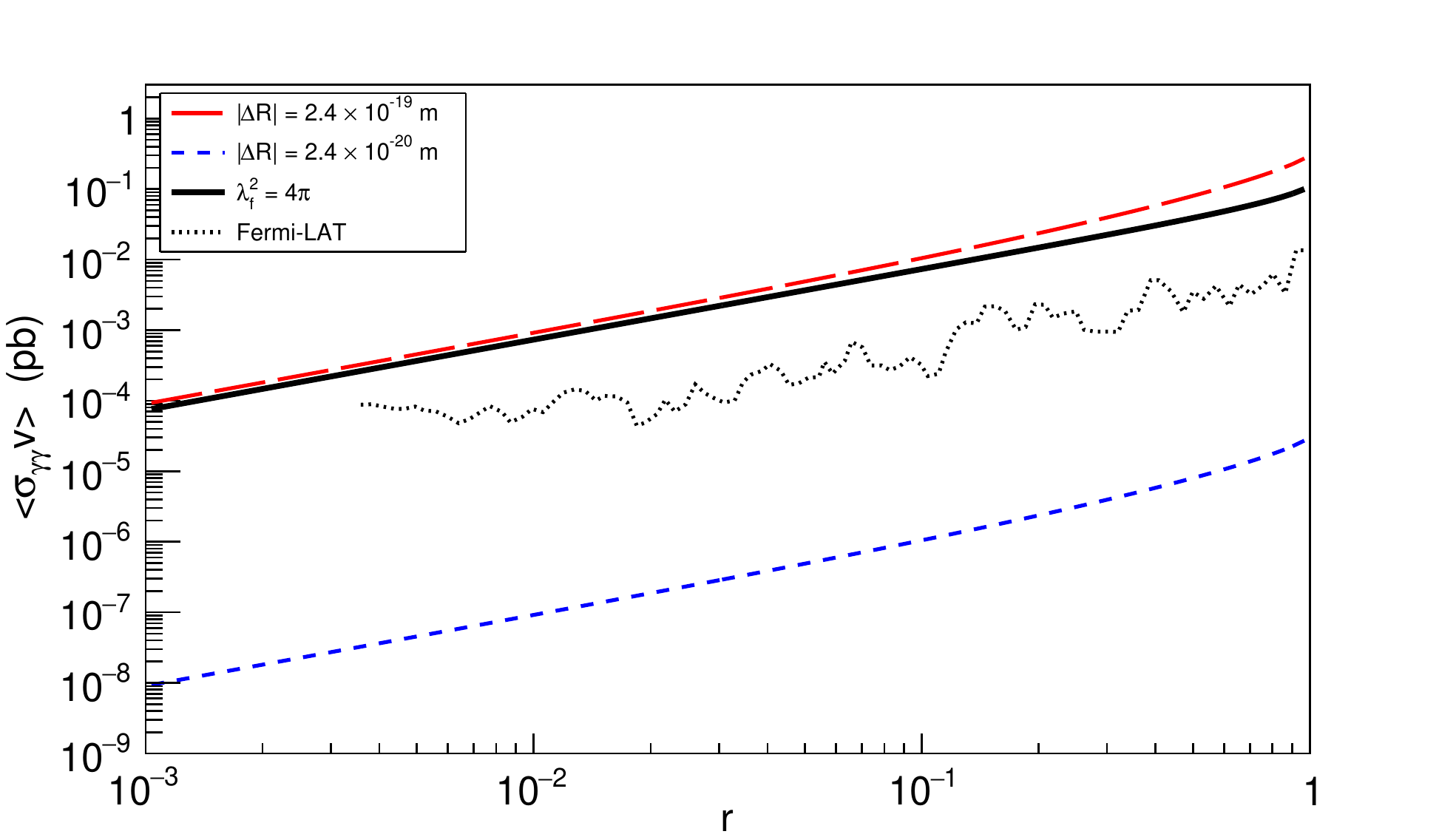}\hspace{0.5cm}
\caption{\label{fig:gg} Bounds on $\langle \sigma_{\bar \gamma \gamma} v \rangle$ (in pb) arising
from measurements of $\Delta R$, where $f=\mu$ and $m_{\tilde f}=300\gev$.
The red long-dashed contour corresponds to the case
where $\Delta R = 2.4 \times 10^{-19}~\m$ (the current limit reported by LEP~\cite{Acciarri:2000uh}), and the
blue short-dashed contour corresponds to the case where $\Delta R = 2.4 \times 10^{-20}~\m$.  The region above the curves
is excluded, if there are no other corrections arising from new physics unconnected to the model we have discussed.  Also shown is the
perturbativity limit (black solid), and observational bounds from Fermi-LAT (black dotted)~\cite{Ackermann:2015lka}.}
\end{figure}

\begin{figure}[ht]
\centering
\includegraphics[width=0.8\textwidth, keepaspectratio]{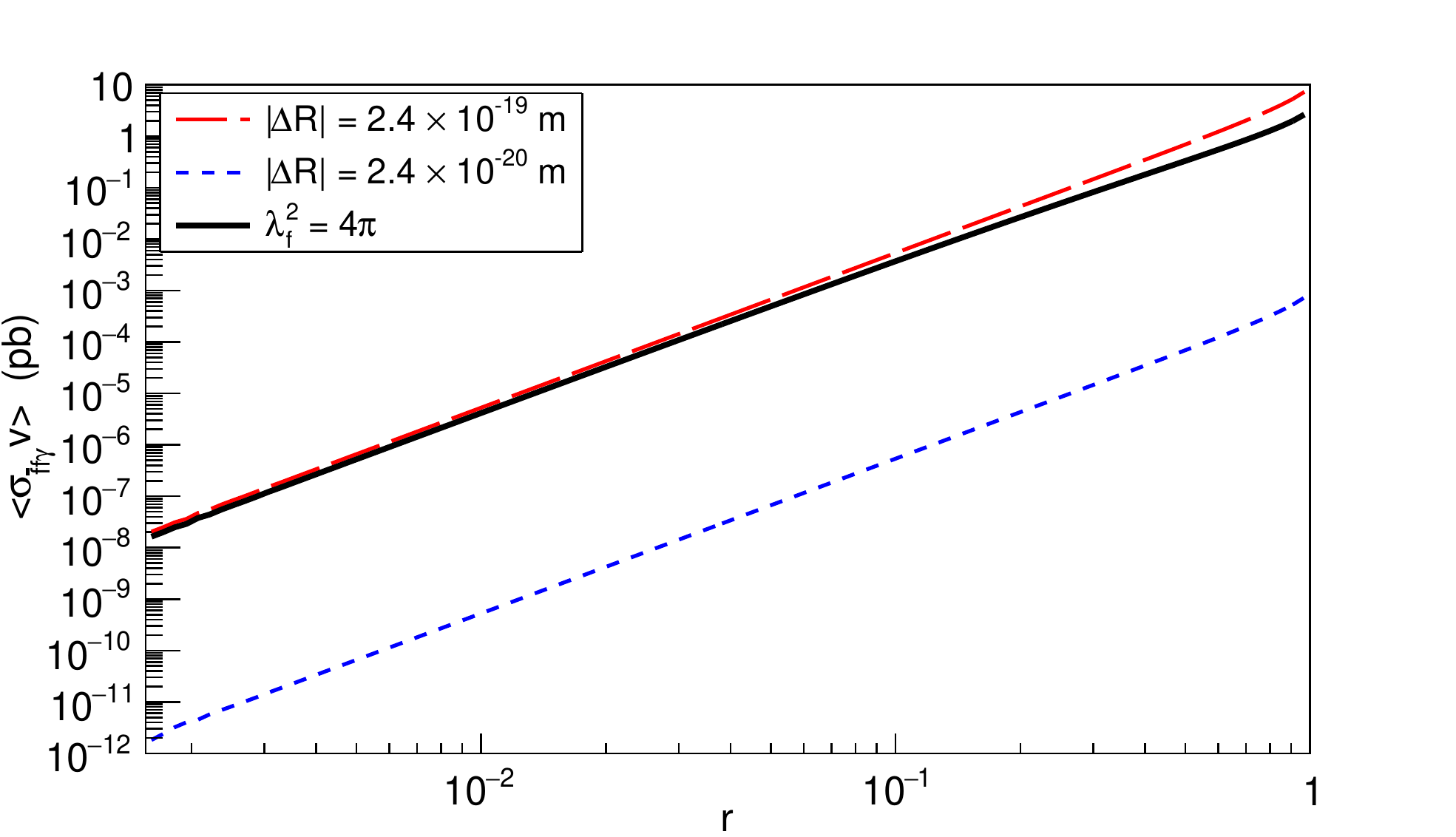}\hspace{0.5cm}
\caption{\label{fig:ffg}
Bounds on $\langle \sigma_{\bar f f \gamma} v \rangle$ (in pb) arising
from measurements of $\Delta R$, where $f=\mu$
and $m_{\tilde f}=300\gev$.
The red long-dashed contour corresponds to the case
where $\Delta R = 2.4 \times 10^{-19}~\m$ (the current limit reported by LEP~\cite{Acciarri:2000uh}), and the
blue short-dashed contour corresponds to the case where $\Delta R = 2.4 \times 10^{-20}~\m$.
The region above the curves
is excluded, if there are no other corrections arising from new physics unconnected to the model we have discussed.  Also shown is the
perturbativity limit (black solid).}
\end{figure}

In Figures~\ref{fig:gg} and~\ref{fig:ffg}, we plot the constraints on $\langle \sigma_{\gamma \gamma} v \rangle$
and $\langle \sigma_{\bar f f \gamma} v \rangle$, respectively, as functions of $r$ ($m_{\tilde f}=300\gev$).
The red long-dashed contour describes the constraints which arise from current LEP bounds on $\Delta R$ for the muon, while
the blue short-dashed contour describes the constraints which would
arise if $\Delta R$ were constrained to be 10 times smaller by future data.
Again, the regions above the contours
are excluded if there are no other sources for corrections to the vertex, and we also plot the largest
cross sections consistent with perturbativity ($\lambda_f^2 =4\pi$).  In
Figure~\ref{fig:gg}, we also plot current observational bounds on $\langle \sigma_{\gamma \gamma} v \rangle$ obtained by Fermi-LAT~\cite{Ackermann:2015lka}.

The process $\chi \chi \rightarrow \gamma \gamma$ can proceed from an $s$-wave initial state, but
$\langle \sigma_{\gamma \gamma} v \rangle$ is suppressed by a factor $\alpha^2 \sim {\cal O}(10^{-4})$.
This cross section will be somewhat larger than $\langle \sigma_{\bar f f} v \rangle$ in the current epoch.
Moreover,
because this annihilation process produces monoenergetic photons,  the signal is much more easily detected above
background.  As a result, this process will be much more important than $\chi \chi \rightarrow \bar f f$ for
indirect detection in the
current epoch.
For $m_{\tilde f} \sim 300\gev$, we find that constraints on the $\chi \chi \rightarrow \gamma \gamma$ annihilation process
arising from collider probes of $\Delta R$ could exceed current constraints from Fermi-LAT by more than three orders of
magnitude (absent fine-tuning), if $\Delta R$ could be constrained at the ${\cal O}(10^{-20}~\m)$ level.

Dark matter annihilation via internal bremsstrahlung is an $s$-wave process, but instead receives a
factor $\alpha r^2$ suppression.  For $r \sim 1$, however, the photon spectrum arising from internal bremsstrahlung
is very hard; the photon spectrum is similar in shape to a line signal, implying that the IB cross section can also be
constrained by line signal searches using Fermi-LAT~\cite{Ackermann:2015lka}.  For $m_{\tilde f} \sim 300\gev$,
constraints on $\langle \sigma_{\bar f f \gamma} v \rangle$ arising from collider probes of $\Delta R$ could surpass
Fermi-LAT constraints on IB in the nearly-degenerate limit by up to two orders of magnitude, if $\Delta R$ can be
constrained at the ${\cal O}(10^{-20}~\m)$ level.

For $r < 1$, however, the photon spectrum will be smooth, and the sensitivity of Fermi-LAT and other
indirect detection experiments will be significantly weaker.  Measurements of $\Delta R$ at the
${\cal O}(10^{-20}~\m)$ level would thus provide an even more dramatic improvement over current indirect detection results.
But for $r \lesssim {\cal O}(0.1)$, however, we see that indirect detection prospects will in any case
tend to be dominated by the process $\chi \chi \rightarrow \gamma \gamma$.

Note, there will also be bremsstrahlung and loop diagram processes in which $W$ and $Z$ bosons
are emitted.  But we have focussed on processes producing hard photons, which are particularly amenable
to detection.

\section{Conclusions}

We have seen that probes of Standard Model fermion form factor corrections can be used to
constrain cross sections for dark matter annihilation in scenarios where dark matter couples
to SM fermions through charged mediators, and where helicity mixing is suppressed.
In particular, the $p$-wave cross section for the process $\chi \chi \rightarrow \bar f f$, and
$s$-wave cross sections for the internal bremsstrahlung process $\chi \chi \rightarrow \bar f f \gamma$ and the
one-loop process $\chi \chi \rightarrow \gamma \gamma$, can be constrained by
probes of the corrections to the fermion $F_1$ form factor, which can be expressed in terms of a parameter
$\Delta R$.
However, it is always possible for other new physics, unrelated
to dark matter, to produce canceling corrections to the form factor, allowing a large annihilation
cross section to be consistent with experimental data.  As such, the bounds we have found do not
correspond to a rigorous limit, but rather demarcate the rough point at which fine-tuning would be
needed in order to retain consistency with the data.

It is worth noting that the annihilation processes $\chi \chi \rightarrow \bar f f \gamma, \gamma \gamma$
can only proceed if dark matter couples to SM fermions through an interaction mediated by a charged
particle.  There will thus always be  a correction to the fermion-photon vertex arising from a one-loop
diagram where dark matter and the charged mediators run in the loop.  Thus, the correlation between the
fermion vertex correction and the $\chi \chi \rightarrow \bar f f \gamma, \gamma \gamma$
annihilation cross sections is robust.

Unfortunately, the constraints on the form factor corrections derivable from LEP data are not precise enough to yield
useful constraints on dark matter annihilation.
But future experiments such as VHEeP~\cite{Caldwell:2016cmw} and ILC could
provide much tighter constraints on the parameter $\Delta R$.
As the annihilation cross sections scale as
$(\Delta R)^4$, even a modest improvement in the constraints on $\Delta R$ would correspond to a dramatic improvement
in constraints on the dark matter annihilation cross section.
For example, a high-luminosity run at VHEeP could yield an an order of magnitude improvement in the measurement of
$\Delta R$~\cite{Caldwell:2016cmw},
resulting in an improvement in bounds on dark matter annihilation of ${\cal O}(10^{-4})$.  Such a dramatic improvement
would result in constraints which essentially exceed all current observational bounds for all of the
channels which we have considered over the entire range of $r$ at $m_{\tilde f} \sim 300\gev$.
High-luminosity $e^+ e^-$ colliders would be ideally
suited for this type of study~\cite{AguilarSaavedra:2001rg}.

Constraining the parameter $\Delta R$ requires one to extract the vertex correction contribution due to
dark matter from collider data.  We have outlined
a strategy for the particular case were dark matter couples to muons $f=\mu$, but this strategy can be
generalized to other choices.  However, there are examples where the extraction of the correction from data can
be more complicated.  For example, if $f=e$, then there will be additional one-loop box diagrams which
contribute to the $e^+ e^- \rightarrow e^+ e^-$ cross section.  In this case, one could attempt to extract
the vertex correction by searching for a similar deviation from lepton universality with a hadronic initial state
($\bar q q \rightarrow e^+ e^- , \bar \mu \mu$), at high energy and/or high luminosity.
However, it is not clear if a similar improvement in the vertex function measurement can be obtained practically.

Finally, one should note that, if future high-energy
experiments do indeed find evidence for a sizeable deviation from the Standard Model prediction for the form
factor, it could be possible to perform a more detailed analysis of the functional form, beyond simply constraining
the first moment.

\section{Acknowledgement}

We are grateful to Tom Browder, Bhaskar Dutta, Danny Marfatia, Xerxes Tata and Sven Vahsen for useful discussions.
JK is supported in part by NSF CAREER grant PHY-1250573.



\end{document}